\documentclass[final]{aa}  
\nolinenumbers

\usepackage{graphicx}

\usepackage{txfonts}

\begin{document} 

\makeatletter
\@ifundefined{if@LN}{}{%
  \global\let\if@LN\iffalse
}
\@ifundefined{c@internallinenumbers}{}{%
  \setcounter{internallinenumbers}{0}%
}
\let\linenumbers\relax
\let\nolinenumbers\relax
\let\pagewiselinenumbers\relax
\let\runninglinenumbers\relax
\makeatother

   \title{Formation and nature of ``Huntsman'' binary pulsars }

   \author{O.G. Benvenuto \inst{1},
           M.A. De Vito \inst{1}, 
           M. Echeveste\inst{2}, 
           M.L. Novarino\inst{1},
           N.D. Pires\inst{3},
           L.M. de S\'a\inst{3},
       and J.E. Horvath\inst{3}
         \fnmsep
     }

\institute{Instituto de Astrofísica de La Plata, IALP, CCT-CONICET-UNLP, Argentina and Facultad de Ciencias
   Astronómicas y Geofísicas de La Plata, Paseo del Bosque S/N, (1900) La Plata, Argentina.    \email{obenvenu@fcaglp.unlp.edu.ar}     
         \and
         INAF - Osservatorio Astrofisico di Arcetri, Largo E. Fermi 5, I-50125 Firenze, Italy
         \and
        Universidade de S\~ao Paulo, Instituto de Astronomia, Geof{\'\i}sica e Ci\^encias Atmosf\'ericas,
              R. do Mat\~ao, 1226, Cidade Universit\'aria, 05508-090 S\~ao Paulo SP, Brazil
             }

   \date{}

  \abstract
   {``Spider'' systems are a class of close binaries in which a neutron star first accretes from a normal companion, and later ablates it in some cases. New observations have expanded this category, with the addition of a  ``Huntsman'' group, tentatively linked to a short donor phase along the red bump along the secondary evolutionary track.}
   {We present here explicit evolutionary tracks that support the Huntsman nature recently suggested, and also discuss how the whole class of spiders emerge from the full consideration of irradiation and ablating winds. We address the irradiation feedback (IFB) effects and the hydrogen-shell burning detachment (HSBD) simultaneously, and show that they act independently and do not interfere with each other, supporting a physical picture of the Huntsman group.}
   {We employed our binary evolution code to compute a suite of binary systems formed by a donor star together with a neutron star for different initial orbital periods, for the case of solar composition and also for $Z=10^{-3}$. Although many models do not consider IFB, we also present the evolution with IFB for one system as an example.}
   {We found that the recently suggested association of Huntsman pulsar with the evolutionary stage where (as consequence of the dynamics of HSBD) the system remains detached for a few million years is truly plausible. However, this feature alone is unable to account for the occurrence of the Redback spider pulsars. Meanwhile, models including IFB, with pulsed mass transfer, display detachment episodes that can be naturally associated with the Redback stage. Irradiation feedback does not preclude or modify HSBD and in fact, the latter were implicit in our earlier calculations, but not addressed explicitly. That is, Huntsman systems were already present as an ``implicit prediction'' in these former works.}
   {We conclude that Huntsman is an expected stage of these spider systems under quite general conditions. This is another step towards a unified picture of spider pulsars as a group. }

\titlerunning{``Huntsman'' spider binary systems}
\authorrunning{O.G. Benvenuto et al.}
\maketitle

\section{Introduction}

It is currently accepted that millisecond pulsars (MSPs) are the product of low-mass X-ray binaries
(LMXBs) or intermediate mass X-ray binaries (IMXBs) when the pulsar passes through a recycling process,
in which accretes matter and angular momentum from its companion star, and is spun up to periods of the
order of milliseconds (i.e., MSPs) \citep{Alpar1982}. An intriguing group of eclipsing MSPs was
identified \citep{Fruchter1988}, which called for an explanation.

More recently, in the last decade or so, the Fermi $\gamma$-ray Space Telescope discoveries have
significantly increased the number of these eclipsing radio MSP systems, with orbital periods of
$P_{orb} < 24$~hours, in the Galactic field. This enabled a classification in these systems based on the
mass of their companion star $M_c$: some of them termed ``Redbacks'' (RBs, $0.1\lesssim
M_c/M_{\odot}\lesssim 0.4$) and others (in which evidence of ablation of the companion was found
\citealt{Fruchter1988}) named ``Black Widows'' (BWs, $0.1\lesssim M_c/M_{\odot}$) \citep{Roberts2013}.
Because of the BW name given after the original discovery, and the RB denomination was  suggested by
Australians for their similarity, the combination of both groups is now known as ``spiders". 

Some ideas have been suggested to explain the spider systems and their possible conceptual unification.
\citet{Chen2013} argued that irradiation by a strong and isotropic wind emitted by the pulsar would form
RBs or BWs, based on how strong the wind really is. Therefore, they concluded that RBs do not evolve
into BWs. Alternatively, \citet{Benve2014} (BDVH14) concluded that for reproducing the spiders' orbital
parameters, X-ray irradiation feedBack (IFB) due to the photons produced by the accretion near the NS is
necessary. They also found that some RBs can be progenitors of BWs. A similar conclusion was recently
reached by \citet{Misra2025}, who were also able to reproduce the two newly discovered spider systems:
``Tidarrens'' \citep{Romani2016} and ``Huntsman'' \citep{Strader2015}. The Tidarrens are, in fact, a
subclass of BWs, but with $P_{orb} < 2$~hrs. Just three of those systems have been detected (see
\citealt{Pletsch2012}; \citealt{Romani2012}; and \citealt{Romani2014}), and the spectra suggest they are
hydrogen-poor, a feature acknowledged by  \citet{Kong2014} for 2FGL J1653.6-0159, and also studied by
\citet{Romani2014}.

At first glance, the Huntsman systems are quite similar to RBs, but their orbital periods are one order
of magnitude larger, and harbor fully recycled pulsars. Just one system is confirmed as a Huntsman
(PSR~J1417-4402; \citealt{Strader2015}), with its spin determined through observations performed with
the  CSIRO Parkes telescope, as 2.66~ms \citep{camilo2016}, and where the distance determination is
still in some doubt, but confirmed to be $\geq 4\ kpc$ with GAIA parallax \citep{Strader2015}. The other
reported system is a strong candidate (PSR~J1947-1120; \citealt{Strader2019}), and new searches should
enlarge the class. 

The Huntsman systems are thought to reach Roche-Lobe OverFlow (RLOF) when the companion star has
exhausted the hydrogen core. Recently, \cite{Strader2025} suggested that those systems originate when
the companion star has reached the RLOF condition in the ``red Bump'' region of the red giant branch,
with initial orbital periods between 2.6~d and 7.0~d. \citet{Misra2025} found that Huntsman pulsars are
reproduced regardless of how strong the pulsar wind is, but the efficiency of the mass accretion during
RLOF is important to reproduce the spin found in the only confirmed system PSR~J1417-4402. This
highlights the importance of considering a broad range of initial orbital periods $P_{orb,i}$ when
studying spider systems.

These findings suggest that the unification of all spider systems would require the implementation of
additional physics, such as IFB and pulsar wind evaporation, to reproduce the positions in the $P_{orb}$
versus $M_{2}$ plane (where $M_{2}$ is the mass of the companion, donor star) and their full
evolutionary tracks. The consideration of a broad range of orbital parameters is also necessary for
reproducing, among others, the Huntsman systems. 

We shall show in this work how Huntsman systems arise by performing explicit binary evolution
calculations. The minimum and maximum periods expected from theory are found assuming the standard
magnetic braking, for both solar metallicity and a low metallicity $Z= 10^{-3}$, which bracket future
systems that have yet to be identified. The behavior of the radius is also given explicitly. In fact,
the evolutionary stage suggested to be associated with Huntsman pulsars was already present in our first
calculations (i.e., BDVH14), although that stage was not specifically analyzed in that paper.

The remainder of this letter is organized as follows: In Section~\ref{sec:huntsman} we present our
evolutionary calculations. Finally, in Section~\ref{sec:disconclu} we discuss the meaning of these
results and give some concluding remarks.

\section{The Huntsman state along the binary evolution} \label{sec:huntsman}

We have employed the code constructed and described in \citet{BDV} to compute the evolution of a suite
of systems formed by a donor star of initial mass $M_{2,i}= 1.25~M_{\odot}$, together with a NS
companion of $M_{NS,i}=1.3~M_{\odot}$ with $P_{orb,i}$ logarithmically evenly spaced with steps of 20\%.
We assumed a conservative mass transfer; considering, as usual, that the NS can accrete matter up to the
Eddington rate: $\dot{M}_{Edd}= 2 \times 10^{-8}\ (M/M_{\odot})\ M_{\odot}\ yr^{-1}$. The orbital
evolution was computed from the ZAMS. For this purpose, we have assumed the same physics as described in
\citet{Maite2024}. Here we considered the standard magnetic braking prescription \citep{Verbunt1981}.
For solar metallicity models we considered $0.69\ d \leq P_{orb,i} \leq 18.48\ d$ whereas for $Z=
10^{-3}$ we employed $0.40\ d \leq P_{orb,i} \leq 46.00\ d$. In these calculations we have not
considered IFB (see below).

We show in Fig.~\ref{Fig:masa_vs_periodo} the evolution of these systems in the $P_{orb}$~versus~$M_{2}$
plane. As is well known, models without IFB undergo a long-standing mass transfer episode on a timescale
of megayears or even gigayears depending on the initial orbital period. The first episode in which the
star detaches from its lobe is due to the stage in which the H-shell burning reaches a step-like
discontinuity in the H-profile left by the very deep outer convective zone. The H-shell burning
detachment will be referred to as "HSBD" in the rest of this work. The star reacts by detaching from its
lobe for a timescale of a few megayears. In the case of isolated stars, this behavior (found long ago
by, e.g., \citealt{Thomas1967}) is connected with the occurrence of a red bump
\citep{2015MNRAS.453..666C} seen in the Hertzsprung~-~Russell diagram (HRD) of globular \citep{Globular}
and open clusters \citep{Abierto}. After detachment, the advancement of the H-shell burning forces the
star to swell substantially to reestablish the RLOF. During detachment, $P_{orb}$ and $M_{2}$ remain
almost unchanged. Although there is gravitational radiation, and maybe some wind mass loss associated
with this stage, both are minor effects. Therefore, there is a well-defined point-like region in the
$P_{orb}$~versus~$M_{2}$ plane, at which detachment occurs. This is shown in
Fig.~\ref{Fig:masa_vs_periodo}, where the heavy blue (pink) dots on each track, corresponding to solar
composition ($Z=10^{-3}$) mark the place where HSBD occurs. It is clear that the two candidates are
located in the galactic plane, and feature a near-solar metallicity. However, since the red bump feature
is expected for a wide range of metallicities, we have added results corresponding to them, because we
do expect this kind of spider pulsar will soon be discovered among objects with these abundances. Thus,
the low-metallicity results have been included as a kind of prediction of the conditions in which we
expect them to be detected. For completeness, in Fig.~\ref{Fig:HR_MB0} we present the HRD corresponding
to the case of solar composition models.

In Tables~\ref{tab:detached}-\ref{tab:detached_Zbajo} we present the main characteristics of the
detached phase of these models. In those tables, $R_{2}$ is the radius of the companion donor star,
whereas $R_{L}$ is the radius of a sphere with a volume equal to the one of its Roche lobe. Notice that,
compared to solar composition models, low-metallicity ones undergo HBSD at higher $P_{orb}$ (starting
from higher  $P_{orb,i}$) and appreciably higher luminosities. This indicates that the H-shell burning
is more intense, making the detachment stage appreciably shorter. Nevertheless, the masses of the
components and filling factors are similar for both compositions considered here.

We should stress that there is a well-defined range of $P_{orb,i}$ values for which HSBD is expected to
occur. In this suite of calculations, for solar composition and $Z=10^{-3}$, it happens if $1~d \leq
P_{orb,i} \leq 12.83~d$ and $1.44~d \leq P_{orb,i} \leq 26.62~d$ respectively. For somewhat lower
$P_{orb,i}$ values, the H-core is not exhausted before RLOF. For higher $P_{orb,i}$ values, the H-shell
burning reaches the H-step profile before RLOF. It should be remembered that these numbers change
slightly with the exact physical ingredients entering the evolution code. The full emerging picture will
be given in a forthcoming work, that will also explore the issue of different initial NS masses and
other related effects.

\begin{table*}
 \centering
 \caption{Main characteristics of the detached stage of  systems with a solar composition donor star of $M_{2,i}=1.25~M_{\odot}$ and a $M_{NS,i}= 1.3~M_{\odot}$. }
 \begin{tabular}{ccccccccc} \hline \hline
  $P_{orb,i}/d$ &
  $T/Gyr$   &
  $\Delta t/Myr$ &
  $M_{2}/M_{\odot}$ &
  $M_{NS}/M_{\odot}$ &
  $Log_{10}(L/L_{\odot})$ &
  $Log_{10}(T_{eff}/K)$ &
  $P_{orb}/d$ &
  $R_{2}/R_{L}$ \\ \hline
  1.00 & 6.210 & 200 & 0.2170 & 2.1016 & 0.4691 & 3.7070 &  3.2396 & 0.868 \\
  1.20 & 5.703 & 108 & 0.2515 & 2.0773 & 0.6129 & 3.6713 &  4.6738 & 0.898 \\
  1.44 & 5.561 &  93 & 0.2677 & 2.0649 & 0.6674 & 3.6644 &  5.2709 & 0.897 \\
  1.72 & 5.479 &  82 & 0.2851 & 2.0518 & 0.7129 & 3.6589 &  5.7044 & 0.901 \\
  2.07 & 5.366 &  69 & 0.3139 & 2.0292 & 0.7942 & 3.6525 &  6.5564 & 0.900 \\
  2.48 & 5.312 &  59 & 0.3554 & 1.9581 & 0.8473 & 3.6488 &  7.0681 & 0.889 \\
  2.98 & 5.278 &  48 & 0.4050 & 1.8970 & 0.9167 & 3.6465 &  7.5675 & 0.890 \\
  3.58 & 5.256 &  42 & 0.4511 & 1.8432 & 0.9970 & 3.6451 &  8.6100 & 0.869 \\
  4.29 & 5.255 &  35 & 0.5067 & 1.8009 & 1.0704 & 3.6445 &  9.1894 & 0.873 \\
  5.15 & 5.253 &  29 & 0.5658 & 1.7376 & 1.1533 & 3.6441 &  9.9515 & 0.878 \\
  6.19 & 5.256 &  24 & 0.6212 & 1.6759 & 1.2381 & 3.6432 & 10.9982 & 0.881 \\
  7.43 & 5.260 &  20 & 0.6598 & 1.6213 & 1.3349 & 3.6404 & 12.8373 & 0.880 \\
  8.91 & 5.263 &  16 & 0.7203 & 1.5524 & 1.4098 & 3.6397 & 14.1156 & 0.875 \\
 10.69 & 5.268 &  13 & 0.7857 & 1.4841 & 1.4915 & 3.6385 & 15.5098 & 0.878 \\
 12.83 & 5.270 &  12 & 0.9633 & 1.3953 & 1.5519 & 3.6439 & 14.2132 & 0.900 \\
 \hline \hline
 \end{tabular}
\label{tab:detached}
\tablefoot{The first column gives the initial orbital period. The second column gives the age of the
system in the middle of the detached stage. The third column gives the time spent in the detached state.
The fourth and fifth columns present the masses of the donor and NS at the Huntsman regime,
respectively. The rest of the columns give the luminosity, the effective temperature, the orbital
period, and the filling factor, all of them evaluated in the middle of the detached stage. }
\end{table*}

\begin{table*}
 \centering
 \caption{Same as Table~\ref{tab:detached}, but for the case of a low metallicity $Z= 10^{-3}$. }
 \begin{tabular}{ccccccccc} \hline \hline
  $P_{orb,i}/d$ &
  $T/Gyr$   &
  $\Delta t/Myr$ &
  $M_{2}/M_{\odot}$ &
  $M_{NS}/M_{\odot}$ &
  $Log_{10}(L/L_{\odot})$ &
  $Log_{10}(T_{eff}/K)$ &
  $P_{orb}/d$ &
  $R_{2}/R_{L}$ \\ \hline
 1.44 & 2.577 & 1.2 & 0.2738 &   1.9452 &   1.5687 &   3.7139 &  17.1142 &   0.912 \\
 1.72 & 2.567 & 2.8 & 0.2812 &   1.9792 &   1.5897 &   3.7068 &  18.5325 &   0.907 \\
 2.07 & 2.560 & 6.0 & 0.2898 &   2.0176 &   1.5993 &   3.7023 &  19.7900 &   0.887 \\
 2.48 & 2.551 & 6.6 & 0.3004 &   2.0076 &   1.6284 &   3.6952 &  22.1774 &   0.869 \\
 2.98 & 2.547 & 6.5 & 0.3073 &   1.9551 &   1.6510 &   3.6911 &  23.4852 &   0.868 \\
 3.58 & 2.544 & 6.6 & 0.3209 &   1.9110 &   1.6784 &   3.6852 &  25.2100 &   0.866 \\
 4.29 & 2.542 & 6.4 & 0.3394 &   1.8761 &   1.7128 &   3.6800 &  27.1441 &   0.862 \\
 5.15 & 2.542 & 6.1 & 0.3636 &   1.8410 &   1.7531 &   3.6753 &  28.8210 &   0.867 \\
 6.19 & 2.541 & 5.6 & 0.3914 &   1.7944 &   1.7990 &   3.6717 &  30.6458 &   0.870 \\
 7.43 & 2.541 & 5.0 & 0.4232 &   1.7421 &   1.8486 &   3.6687 &  32.4989 &   0.875 \\
 8.91 & 2.542 & 4.8 & 0.4583 &   1.6821 &   1.9039 &   3.6660 &  34.5665 &   0.882 \\
10.69 & 2.543 & 3.9 & 0.5004 &   1.6127 &   1.9549 &   3.6644 &  36.5779 &   0.880 \\
12.83 & 2.544 & 3.5 & 0.5475 &   1.5474 &   2.0090 &   3.6629 &  38.4641 &   0.884 \\
15.40 & 2.545 & 3.0 & 0.5964 &   1.4923 &   2.0643 &   3.6616 &  40.6381 &   0.886 \\
18.48 & 2.546 & 2.6 & 0.6437 &   1.4481 &   2.1258 &   3.6600 &  43.4772 &   0.891 \\
22.18 & 2.548 & 2.4 & 0.7040 &   1.4102 &   2.1851 &   3.6589 &  45.1874 &   0.904 \\
26.62 & 2.548 & 2.1 & 0.8125 &   1.3710 &   2.2145 &   3.6620 &  43.0515 &   0.902 \\
 \hline \hline
 \end{tabular}
\label{tab:detached_Zbajo}
\end{table*}

\begin{figure}
\centerline{\includegraphics[width=0.40\textwidth,angle=270]{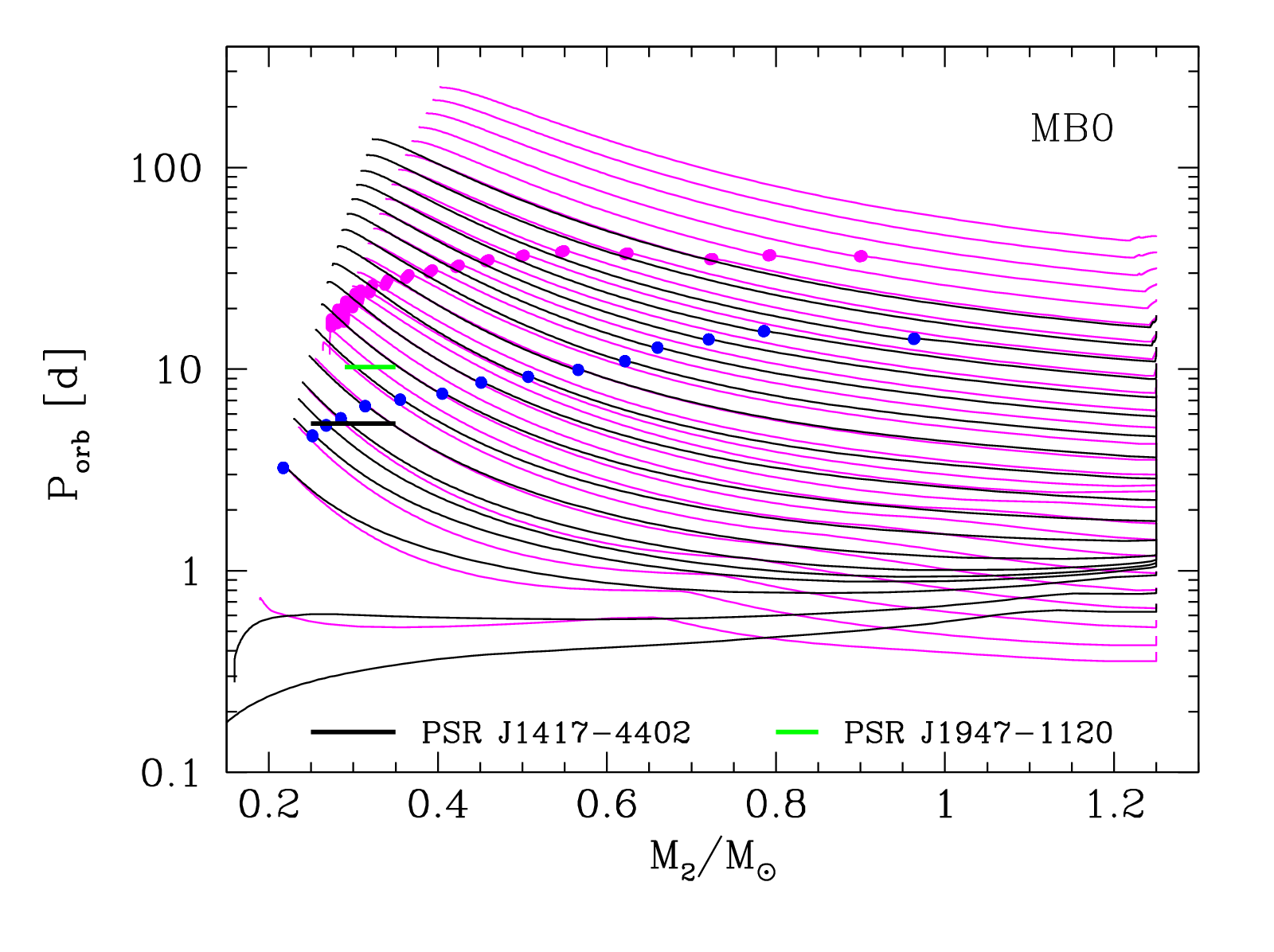}}
\caption{The orbital period as a function of the donor mass for a suite of systems formed by a donor star of 1.25~$M_{\odot}$, a NS of 1.3~$M_{\odot}$ with $P_{orb,i}$ logarithmically evenly spaced with steps of 20\%. The orbital evolution was computed from the ZAMS. Black (magenta) lines denote results corresponding to solar metallicity ($Z=10^{-3}$). The detached stages, in which Huntsman pulsars are expected, are denoted in blue for solar metallicity and in pink for $Z=10^{-3}$. Horizontal bars correspond to the measurements of the system PSR~J1947-1120 (green) and PSR~J1417-4402 (black). MB0 denotes the standard magnetic braking prescription.}
\label{Fig:masa_vs_periodo}
\end{figure}

\begin{figure}
\centerline{\includegraphics[width=0.40\textwidth,angle=270]{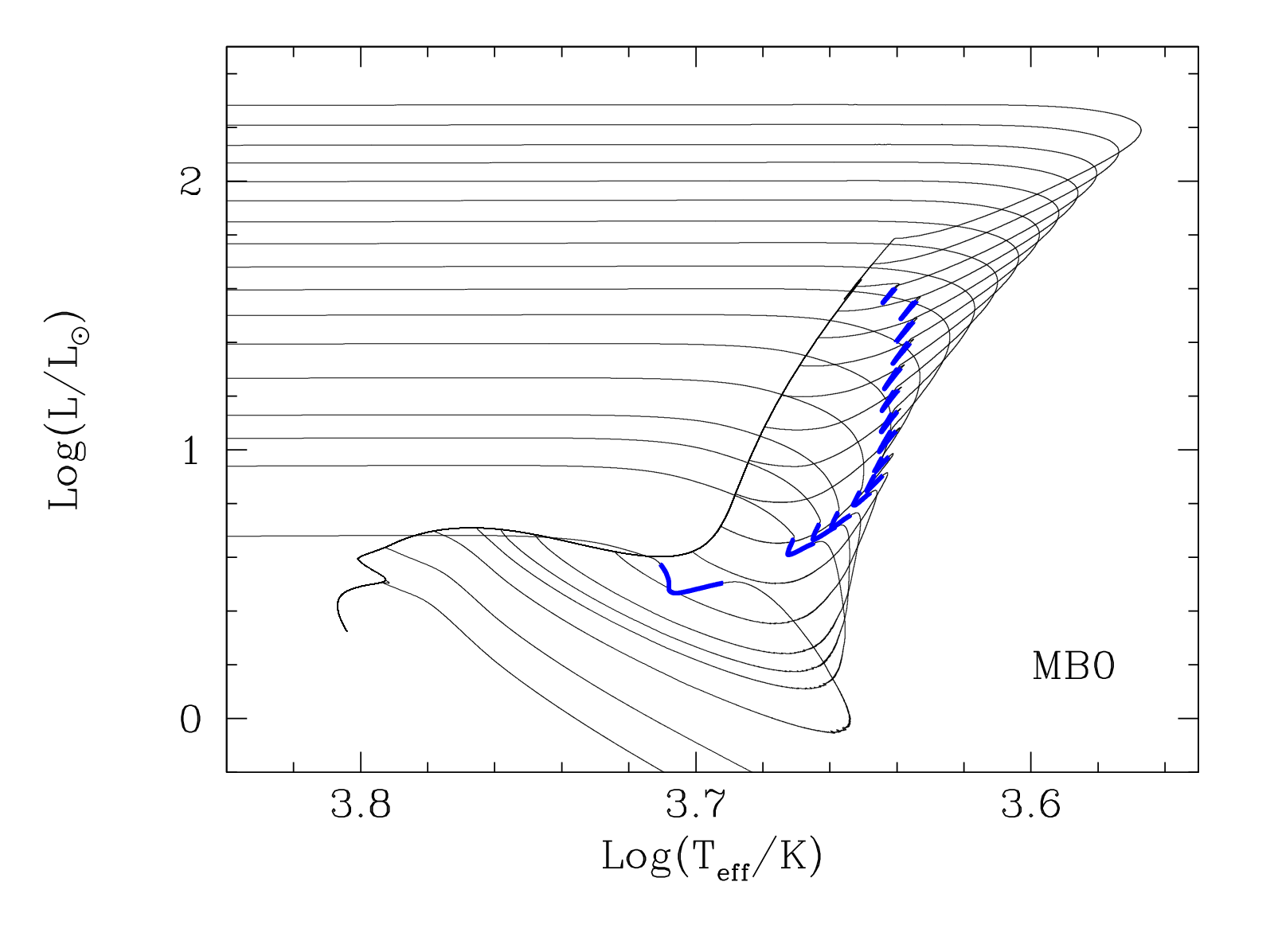}}
\caption{HR diagram showing the trajectories and the intervals where the Huntsman states 
occur (blue). The first initial period that allows the mass transfer (the so-called case B) is clearly
seen on the bottom, producing the ample and slanted ``V'' track is $P_{orb, i} = 1 \, d$, while all
shorter periods end in downward tracks. It is important to remind the reader that this is somewhat
sensitive to the treatment of the physics and small variations may occur for different choices of the
braking and other features (see, e.g., \citealt{Thomas1967, Mis}).}
\label{Fig:HR_MB0}
\end{figure}

\begin{figure}
\centerline{\includegraphics[width=0.40\textwidth,angle=270]{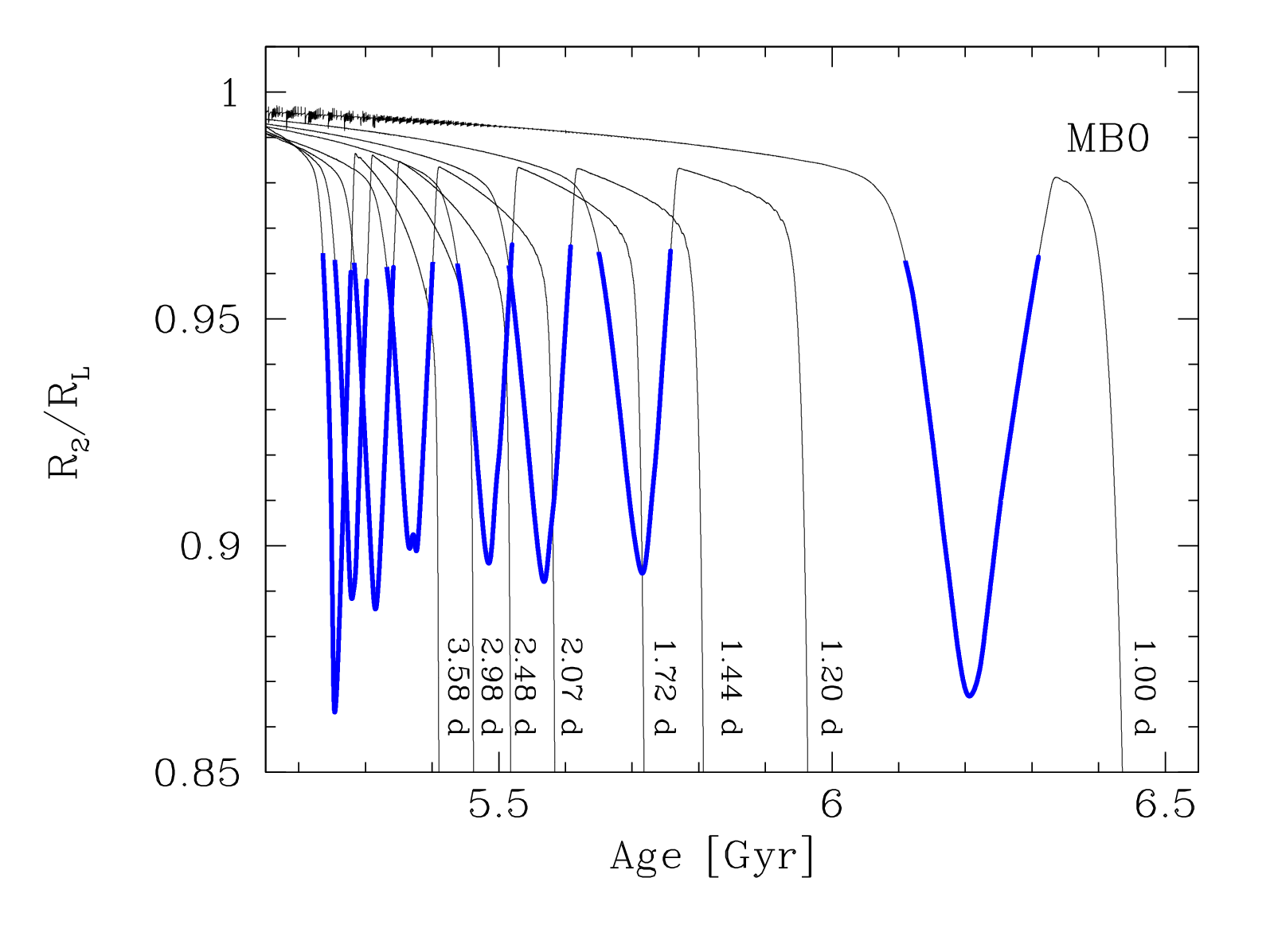}}
\caption{$R_{2}/R_{L}$ filling factor for some of the solar composition models presented in Fig.~\ref{Fig:masa_vs_periodo} as a function of age. Models with larger $P_{orb,i}$ that undergo HSBD are not included since they behave very similar to the case of $P_{orb,i}= 3.58$~d. Labels on each curve correspond to $P_{orb,i}$. The fractions marked with heavy blue lines correspond to conditions in which the system is detached, allowing for the detection of the pulsar companion.}
\label{Fig:detached_vs_tiempo}
\end{figure}

In Fig.~\ref{Fig:detached_vs_tiempo} we show the Roche lobe filling factor defined as $R_{2}/R_{L}$ as a
function of time for solar composition models. There, we have not included all those undergoing HBSD
since for those with $3.58\ d \leq P_{orb,i}\leq 12.83\ d$, curves are very similar. Interestingly, the
values of the calculated filling factors are high, in the range of $0.868\ -\ 0.901$ (see
\citet{Filling_Factors}, especially their Table~2).

\begin{figure}
\centerline{\includegraphics[width=0.40\textwidth,angle=270]{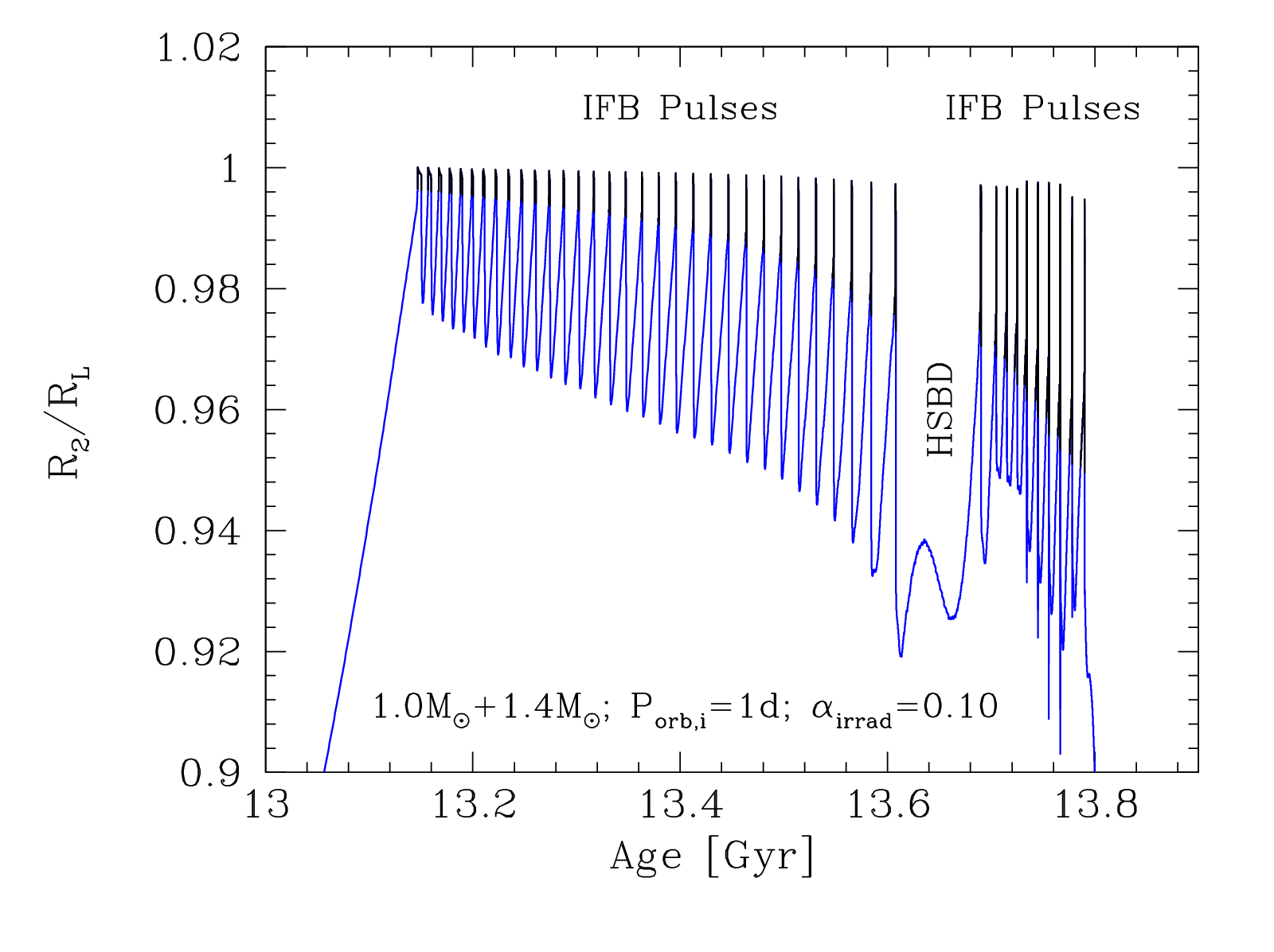}}
\caption{$R_{2}/R_{L}$ filling factor for a system formed by a solar composition donor star of $1M_{\odot}$, a NS of $1.4M_{\odot}$ with $P_{inic,i}=1\ d$. We considered moderate IFB assuming $\alpha_{irr}= 0.10$. Blue lines represent detached conditions whereas black lines depict stages with a mass transfer rate $\dot{M} \geq 10^{-11}\ M_{\odot}\ y^{-1}$. IFB-driven pulsed mass transfer and HSBD occur in the model.}
\label{Fig:El_Monio}
\end{figure}

An important issue is whether models with IFB-driven pulsed mass transfer also undergo HSBD. To address
this issue, we have computed a set of models with IFB similar to those presented in BDVH14. To include
IFB, we assumed the \citet{hameury-ritter} treatment: the irradiation flux $F_{irr}$ is given by
$F_{irr}= \alpha_{irr} L_{irr}/(4\pi a^{2})$ where $\alpha_{irr}\leq 1$ is the fraction of accretion
luminosity $L_{irr}$ that effectively participates in the donor irradiation, and $a$ is the orbital
semiaxis. Furthermore, $L_{irr}= G \dot{M}_{NS} M_{NS}/R_{NS}$ where $G$ is the gravitational constant
and  $\dot{M}_{NS}$, $M_{NS}$, and $R_{NS}$ are the accretion rate, the mass and the radius of the NS
companion, respectively. Among the irradiated models, we have selected the case of a system formed by a
donor star of  $M_{2,i}=1\ M_{\odot}$, a NS of $M_{NS,i}= 1.4\ M_{\odot}$ with $P_{orb,i}=1\ d$ and
moderate IFB assuming $\alpha_{irr}= 0.1$. In Fig.~\ref{Fig:El_Monio} we present the filling factor as a
function of time for the entire time interval in which the system undergoes mass transfer. It is seen
that pulsed mass transfer occurs before and also after HSBD. This is an expected situation, since HSBD
is nuclear burning-driven, acting on very deep layers, whereas IFB is essentially a surface
phenomenon.  

\section{Discussion and conclusions}\label{sec:disconclu}

To account for the very existence of BWs and RBs, it has been suggested in previous works
(\citealt{Benve2014}, BDVH14) that models have to include IFB and/or evaporation. Both phenomena act on
the surface of the donor star. On the contrary, HSBD is associated with the deep interior of the star,
and does not depend on the former, which acts quite independently in these systems.

An important feature is that the two detected Huntsman systems have relatively short periods (5.374~d
and 10.265~d for  PSR~J1417-4402 and PSR~J1947-1120, respectively). This automatically means that their
companion stars belong to the less massive end of the donor mass range ($M_{2} \lesssim 0.5\ M_{\odot}$)
which allows a Huntsman state. The shorter period system is particularly restrictive (see
Fig.~\ref{Fig:masa_vs_periodo}), although a lower metallicity and a different initial donor mass would
improve the agreement, as we shall show in detail elsewhere.

In the calculations presented in \citet{Benve2015} we have considered IFB that leads to pulsed mass
transfer. Depending on the strength of irradiation, when detached, models have values of $0.92 \leq
R_{2}/R_{L} \leq 0.99$ (see Fig.~7 of \citealt{Benve2015}). IFB does not quench HSBD. Indeed, IFB-driven
pulsed mass transfer and HSBD occur in the same model star as shown above, in Fig.~\ref{Fig:El_Monio}.  

Alternatively, \citet{Chen2013} have suggested that BWs and RBs are due to different evaporation regimes
and follow separate evolutionary tracks. In that picture, BWs (RBs) undergo moderate (severe)
evaporation. Since the Roche lobe filling factor values attained in these models are far lower than the
filling factor during HSBD \citep{Filling_Factors}, model detachment is led by evaporation and precludes
the occurrence of HSBD.

Thus, we conclude that the identification of the Huntsman pulsars with HSBD suggested by
\citet{Strader2025} seems very plausible. However, additional ingredients are still needed to account
for all spider pulsars. Indeed, it seems natural that pulsars with high enough values of $P_{orb,i}$
(although not too high) will undergo HSBD and also pulsed mass transfer due to IFB action. Depending on
the precise value of $P_{orb,i}$, they will evolve through the RB stage, and some will finally reach the
BW stage. If evaporation is not too weak, they will become ``normal'' BWs. If evaporation is weak, the
orbit will remain close to that predicted by standard models that are driven solely by gravitational
radiation and will become Tidarrens. Hunstman systems are, as discussed above, one relatively short
stage expected from evolutionary considerations, emerging from the overall spider family. Thus, this
picture (with ingredients anchored in observational evidence, IFB \citep{irr} and ablation
\citep{Fruchter1988} provides a natural and quite complete view of the evolution of these systems.

\begin{acknowledgements}
OGB is a member of the Carrera del Investigador Científico of the Comisión de Investigaciones Científicas Of the Province of Buenos Aires, Argentina (CIC-PBA). MADV is a member of the Carrera del Investigador Científico of the Consejo Nacional de Investigaciones Científicas y Técnicas (CONICET). LN is a CONICET fellow. This work was supported by the FAPESP Agency (S\~ao Paulo State) under the grant 2024/16892-2 and the CNPq (Federal Government) for the award of a Research Fellowship to JEH. \\
Co-funded by the European Union (ERC-2022-AdG, {\em "StarDance: the non-canonical evolution of stars in clusters"}, Grant Agreement 101093572, PI: E. Pancino). Views and opinions expressed are however those of the author(s) only and do not necessarily reflect those of the European Union or the European Research Council. Neither the European Union nor the granting authority can be held responsible for them.
\end{acknowledgements}

\end{document}